\documentclass[aps,preprint,groupedaddress,showpacs,showkeys,nofootinbib]{revtex4}
\usepackage{graphicx}

\newcommand{\be}{\begin{equation}} 
\newcommand{\ee}{\end{equation}}

\newcommand{\bea}{\begin{eqnarray}} 
\newcommand{\eea}{\end{eqnarray}} 
\newcommand{\eg}{e.g., } 
\newcommand{\ie}{i.e., } 
\newcommand{\OM}{\Omega_m} 
\newcommand{\OO}{\Omega_0}
\newcommand{\OL}{\Omega_{\Lambda}} 

\newcommand{\nn}{\nonumber\\}

\newcommand{\bO}{b_{\Omega}}

\begin{document}
\bibliographystyle{apsrev}
\title{The  Lame$^{\prime}$  Equation for Distance-Redshift in Partially Filled Beam
Friedmann-Lema\^\i tre-Robertson-Walker Cosmology} 
\author{ R.  Kantowski }
\email{kantowski@mail.nhn.ou.edu}
\affiliation{
Department of Physics and Astronomy, University of Oklahoma\\
Norman, OK 73019, USA}
\date{\today}
\begin{abstract} \vskip .2 truein 
The differential equation governing distance-redshift  for partially filled-beam optics in pressure-free
Friedmann-Lema\^\i tre-Robertson-Walker Cosmology (FLRW) is shown to be the Lame$^{\prime}$ equation.
The distance-redshift, $D(z)$, discussed is appropriate for observations in inhomogeneous cosmologies 
for which lensing by masses external to the observing beam are negligible and for which lensing 
by transparent matter within the beam can be approximated by a homogeneous mass density 
expanding with the FLRW background. 
Some solutions of the derived Lame$^{\prime}$ equation are given in terms 
of Weierstrass elliptic integrals. 
A new simplified and useful expression for filled-beam $D(z)$ in 
standard flat FLRW is also given.
\end{abstract}

\keywords{cosmology:  theory -- large-scale structure of universe}
\pacs{98.80.-k}
\maketitle
\newpage
\section{INTRODUCTION TO GRAVITATIONAL OPTICS} \label{intro}
One constant goal of cosmology has been to determine the large scale geometrical 
structure of the Universe, i.e., the Hubble parameter $H_0$ and the mass density parameter $\OM$. 
More recently an additional dynamical parameter $\OL$ (a cosmological constant or vacuum energy term) 
has been added to the ``must know" list and even more exotic forms of matter and associated parameters 
are being speculated about, e.g., quintessence, see \cite{MBMS}.  
One statistically significant constraint on these parameters comes from fitting 
the observed magnitude-redshift of type Ia supernovae to Hubble curves predicted by the FLRW models,
\cite{RAG, SB, PS1}. 
Magnitude-redshift $m(z)$, or equivalent distance-redshift $D(z)$, for FRW (same as FLRW but without 
$\Lambda$) was first given by Mattig \cite{MW}.
In contrast to
the real Universe, FLRW models are  homogeneous and isotropic on all scales.  The first modification
to the theoretical Hubble curves to account for the lack of homogeneity dates back to Zel'dovich \cite{Zel}. This 
modification amounted to correcting for the decreased converging effect of the decreased
mass density in the line of sight between the observer and the distant object. The idea for this 
modification is that: (i) inhomogeneous 
matter exists in the universe at the expense of homogeneous matter, $\rho_I=\rho_0-\rho_H$,
 (ii) the effects of inhomogeneous mass 
should be accounted for by  gravitational lensing. It is this same correction to the 
Hubble curve being discussed here. 

The original geometrical construction by Zel'dovich was for a flat $\Lambda=0$ 
universe in which all 
matter was absent from the observing beam, $\rho_H=0$. He was the first to give a differential equation from 
which a corrected distance-redshift was derived. The cosmological constant, $\Lambda$, was noticeably absent in 
most calculations done in those days; it was considered to be nothing more than a mistake made by Einstein. 
The first derivation of the differential equation for $D(z)$ with $\OM\ne 1$  
was by Dashevskii and Zel'dovich \cite{DZ}  but again only for the empty beam case. 
It was Dashevskii and Slysh \cite{DS} that extended the equation to partially filled observing beams, $\rho_H\ne 0$.
The differential equation evolved as the number of parameters increased  and the types of cosmologies 
changed, see Kantowski \cite{KR} and for some history of solutions see
\cite{KR98}.  

The first attempt to make use of the corrected magnitude-redshift by Bertotti \cite{BB} was met with 
widespread skepticism. The primary objection (which is valid for weak lensing but not for strong) was that the 
mean magnitude of 
a distant object didn't agree with the standard homogeneous results of Mattig \cite{MW}.   
The primary source of the `error' 
was thought to be the perturbative nature of the cosmological model used. In Betrotti \cite{BB} part of the 
homogeneous FRW mass density was removed and the effect of this missing mass density on optical observations 
was accounted for in accordance with the earlier work of Zel'dovich \cite{Zel}. 
Additional point masses were superimposed to account for lensing by  
nearest neighbor galaxy impacts. 
Another more comprehensive numerical calculation by Gunn \cite{GJ} did not contradict this previous work. 

To overcome the suspected approximation error, the author was encouraged to 
develop geometrical optics for a type of universe called Swiss cheese. This universe
is homogeneous FLRW with numerous spherical holes removed and replaced by point masses 
Schwarzschild solutions \cite{ES} or other
spherically symmetric matter \cite{BH}, (see \cite{KA} for a history of these solutions)
appropriately constrained to
satisfy Einstein's gravity theory. Because this model was an exact solution to Einstein's GR, 
questions related to perturbative 
errors disappeared. With the introduction of the complex Swiss cheese model came the necessity of
following waves as they pass from homogeneous FLRW into and through inhomogeneous Schwarzschild
or Bondi and out again and so on. With Swiss cheese it could be demonstrated that 
a wave going through an inhomogeneity
re-emerged with an insignificantly modified redshift-radius relation ($1+z=R_0/R$) and that the 
affine parameter value was likewise unaltered from FRW. These results were necessary, but were
suspect assumptions made when using superposition models. 
Geometrical optics in Swiss cheese could be done only because of 
the earlier work of Sachs \cite{SR} and  Jordan, Ehlers, and Sachs \cite{JES} that 
developed the optical-scalar equations 
for light propagation in any gravity field.

The geometrical optics approximation for light waves is equivalent to a bundle of non-rotating 
rays traveling 
from a source
in the distant past to  an observer here and now. It is the fractional increase in the area $A$
of this beam
that determines the fractional decrease in the luminosity of the object as seen 
by us, $\ell\propto 1/A$. 
What complicates evaluation of the area $A$ of an 
observing beam is the existence of two accompanying properties, the distortion of the beam (measured 
by its shear $\sigma$ or equivalently $\xi\equiv \sigma\, A$ used below) 
and the orientation of the distortion (measured by an angle $\phi$). The dynamics of these three 
parameters are intricately coupled as can be seen in 
the following form  Kantowski \cite{KR} of the optical scalar equations of \cite{SR} and \cite{JES}.
\bea
&&
{{\sqrt{A}}^{\prime\prime}\over \sqrt{A}}+{\xi^2\over A^2}={\cal R},
\nn
&&
\xi'=A{\cal C} \cos(\beta-\phi),
\nn
&&
\phi'\xi=A{\cal C} \sin(\beta-\phi).
\label{Sachs}
\eea
These equations are complicated in their details but are easy to understand in principle. They contain 
three independent variables $A,\xi,$ and $\phi$ that represent properties of a particular geometrical optics 
wave. The other three terms ${\cal R,C,}$ and $\beta$
are driving terms which are properties of the gravitational field in which the wave propagates.
To use these equations first connect the source and observer with a single central light ray,
a null geodesic of the gravity field, parameterized by an affine parameter, i.e., the prime derivatives 
in (\ref{Sachs}). 
You then start  
the three independent wave variables at the source with initial data appropriate for a point source 
in Euclidian geometry,
$A\propto r^2$, $\xi=0$, and $\phi=$ arbitrary. As the bundle of rays propagates through the 
gravity field they are converged ($A$ is decreased) by the Ricci tensor (the negative ${\cal R}$ term)
and are distorted or sheared (i.e., $\xi=\sigma A$ becomes non-zero)  by the Weyl tensor ${\cal C}$ 
(the conformal curvature part of the Riemann tensor). 
The Ricci tensor is determined by the local density of matter within the bundle of rays whereas
the Weyl tensor is determined  by inhomogeneous matter near but external to 
the rays. The orientation of the distortion of the rays $\phi$ is driven  by the
orientation $\beta$ of the inhomogemeities in the Weyl tensor. 

For homogeneous FLRW, ${\cal C}=0$ (which keeps the shear term $\xi=0$), 
${\cal R}= -3/2\OM(1+z)^5$
and the affine parameter 
is related to the redshift by 
\be
{'}\
\equiv -(1+z)^3 \sqrt{1+\OM z+\OL[(1+z)^{-2}-1]} \ {d \ \over d z}\ .
\label{affine} 
\ee
For Swiss cheese with Schwarzschild inhomogeneities, (\ref{Sachs}) reduces to a single linear and 
homogeneous $3^{rd}$-order  differential equation for $A$, provided the light beam interacts with 
enough randomly distributed inhomogeneities  
as it travels from source to observer, \cite{KR}. 
This $3^{rd}$-order equation was an analytic version of \cite{BB} and \cite{GJ} which 
accounted for the decreased mass density in the observing beam and for the 
lensing of galaxies, except it was derived for an exact GR cosmology.
In \cite{DR74} the entire
calculation was extended to include $\Lambda\ne 0$ by simply including the $\OL$ term in (\ref{affine}).

A more universal application of (\ref{Sachs}), as recognized by Dyer and Roeder \cite{DR72}, occurs when 
the beam interacts infrequently with inhomogeneities or when the interactions introduce negligible shear.
In this case $\xi\approx 0$ and $\sqrt{A}$ satisfies a linear homogeneous second order differential 
equation:
\be
{{\sqrt{A}}^{\prime\prime}\over \sqrt{A}}={\cal R}.
\label{Sachs_sigma=0}
\ee
Values for ${\cal R}$ can be computed for exact gravity fields such as for FLRW as given above 
and for Swiss cheese where ${\cal R}$ fluctuates, dropping from the FLRW value to zero in voids.
For Swiss cheese, \cite{KR} and \cite{DR74}, where the driving term ${\cal R}$ can be replaced by its average along the 
light path the area equation becomes,
\be
{{\sqrt{A}}^{\prime\prime}\over \sqrt{A}}= -\frac32\ {\rho_H\over\rho_m}\OM(1+z)^5,
\label{pfb_eqn}
\ee
where $\rho_H/\rho_m$ is the filling fraction (often denoted by $\alpha$) 
and is the transparent fraction of the total matter density that appears in the beam.
Many refer to this equation as the Dyer-Roeder equation, however,  to do so does an
injustice to Zel'dovich who first created and solved the equation, to Sachs whose optical scalar 
equations it is rigorously derived from, and to others who developed and solved the equation
prior to Dyer and Roeder's first publication \cite{DR72}.

Modification of (\ref{affine}) and the value of the Ricci term in (\ref{Sachs_sigma=0}) can also be made to 
account for additional matter contents of the FLRW model such as quintessence, etc.,  
see \cite{LE},\cite{GOA}, \cite{SCPdeR}, \cite{SPS} and \cite{WFGL}.
It should be understood that there are no exact Swiss cheese type models containing pressure 
in which to verify perturbative assumptions such as the affine parameter and the radius-redshift 
remaining as they are in the corresponding homogeneous FLRW model. 

In the next section (\ref{pfb_eqn}) is converted into its Lame$^{\prime}$ form and in the following section 
it is solved to give $D(z)$ for some special cases. One new result (\ref{Dl_nu0FlatHyper}), seems particularly useful, 
that of flat filled-beam FLRW.

\section{THE PARTIALLY FILLED BEAM EQUATION} \label{pfb}
In this section the partially filled beam area-redshift equation 
for pressure-free FLRW
is revealed to be a standard Lam\'e equation when the dependent 
and independent variables are appropriately redefined.

 We first rewrite equation (\ref{pfb_eqn}) using (\ref{affine}) as
\bea 
&&(1+z)^3\sqrt{1+\OM z+\OL[(1+z)^{-2}-1]}\times\nonumber\\ &&\hskip 1 in {d\ \over
dz}(1+z)^3\sqrt{1+\OM z+ \OL[(1+z)^{-2}-1]}\,{d\ \over dz}\sqrt{A(z)}\nonumber\\ &&\hskip 2.0 in
+ {(3+\nu)(2-\nu)\over 4}\OM(1+z)^5\sqrt{A(z)}=0,  
\label{area} 
\eea
where the parameter $\nu$ (the filling parameter) has been introduced to accommodate standard mathematical notation for special 
functions
\bea
{\rho_I\over \rho_m}&=&{\nu(\nu+1)\over 6},\ \ \ \ 0\le\nu\le 2\nn
{\rho_H\over \rho_m}\equiv \alpha&=&1-{\rho_I\over \rho_m}={(3+\nu)(2-\nu)\over 6}.
\eea
Luminosity distance is
related to a particular solution $A(z)$ of this  equation by
\be
D_{\ell}\equiv(1+z_s) \sqrt{{A\big|_0 \over \delta\Omega_s}}.
\label{Dl} 
\ee
$A|_0$ is the observed area of a beam of light that started expanding with inital 
solid angle $\delta\Omega_s$
from a point source 
at redshift $z_s$. It is evaluated by integrating equation (\ref{area}) from the
source $z=z_s$ to the observer $z=0$ with initial data which makes the wave front satisfy
Euclidean geometry when leaving the source (area=radius$^2\times$ solid angle):  

\bea
\sqrt{A}|_s&=&0,\nonumber\\ {d\sqrt{A}\big|_s\over dz}&=& -\sqrt{\delta\Omega_s} {c\over
H_s(1+z_s)}, \label{Aboundary} 
\eea 
where in FLRW the value of the Hubble parameter at $z_s$ is
related to the current value $H_0$ at $z=0$ by 
\be
H_s=H_0(1+z_s)\ \sqrt{1+\OM
z_s+\OL[(1+z_s)^{-2}-1]}.  \label{Hs} 
\ee
In  \cite{KR98} equation (\ref{area}) was transformed 
into the form of a standard Heun equation by changing the independent variable from $z$ to $y$ and the dependent
variable from $\sqrt{A(z)}$ to $h$.

\subsection{$\OO\equiv \OM+\OL\ne 1$, The Non-Flat FLRW Models} \label{OOne1}

For the non-flat cases the required change of variables is 

\bea 
y&=&{\OM\over 1-\OO}(1+z),\nonumber\\ 
h&=&(1+z)\sqrt{{A\over\delta\Omega_s}}.
\label{heunvariables} 
\eea
The resulting equation is 
\be
{d^2h\over dy^2} +{\left(1+{3\over 2}y\right)y\over y^3+y^2-\bO}\
{dh\over dy} - { {1\over 4}\nu(\nu+1)y+1\over y^3+y^2-\bO}\ h = 0\ , 
\label{heun} 
\ee
where the constant $\bO$ depends on the two FLRW density parameters $\OM$ and $\OL$,

\be
\bO\equiv -\OM^2\OL/(1-\OO)^3.
\label{bO}
\ee
Equation (\ref{heun}) was recognized as a Heun \cite{HK} equation, (see \cite{RA} for details)
and its solution was given in terms of linear Heun 
functions, $Hl$ by Kantowski \cite{KR98}. Sereno et al. \cite{SCPdeR} found that it remains Heun if the 
cosmological constant is replaced by domain wall sources; however,  
when replaced by string networks it reduces to  
hypergeometric. Equations for other types of quintessence sources are yet to be classified. 

Even though the Heun equation is only slightly more 
complicated than the hypergeometric
equation (\eg it has 4 regular singular points rather than 3)
Heun functions are not yet available in standard computer libraries. 
Consequently, such expressions are not yet 
useful for comparison with data. 
Because the exponents of three of the singular points of the 
area equation (in standard Heun form) are 0 and 1/2 
[see Eqn.\,(13) in \cite{KR98}], equation (\ref{heun}) is in fact equivalent to the Lam\'e equation,
as was pointed out at that time.
Details of transforming (\ref{heun}), \ie (\ref{area}) into Lam\'e form will now be given. 
The required change of 
independent variable
is:
\be
u(z)\equiv\int_{\infty}^{p(z)}{dt\over\sqrt{4t^3-t/12-g_3}}\ < 0,
\label{u}
\ee
where integration is along the positive real axis and
\be
p(z)\equiv {\OM\over 4|1-\OO|}(1+z)-{\kappa\over 12}\ > -{\kappa\over 12}.
\label{p}\ee
The constant $g_3$ is given by 
\be
g_3={\kappa\over 3^32^3}(1-b),
\ee
where the constant $b$ is related to $\bO$ of (\ref{bO}) as in \cite{KR98},
\be
b\equiv\frac{27}2\,\bO = -(27/2){\OM^2\OL\over (1-\OO)^3}\ ,
\ee
and the parameter  $\kappa$ is determined by the sign of the 
3-curvature
\be
\kappa\equiv{\OO-1\over |\OO-1|}=\pm1.
\ee
The  three cubics appearing in (\ref{area}), (\ref{heun}), and (\ref{u}) are related by:
\bea
&&4\,p^3-\frac1{12}p-g_3 = -\kappa\frac1{4^2}\left(y^3+y^2-\bO\right)\nonumber\\
&&=\left({\OM\over 4}\right)^2{1\over |1-\OO|^3}
\biggl\{(1+z)^2(1+\OM z)+\OL[1-(1+z)^2]\biggr\},
\eea
and are all positive for Big Bang cosmologies.
The resulting Lam\'e equation\footnote{See section 15.2 of \cite{Erdeliy}. Equation (\ref{Lame}) above is not the most general 
Lam\'e equation. In general it contains two additional arbitrary 
parameters: $H$ rather than $\kappa(1-{\nu(\nu+1)/12})$, and $g_2$  as in (\ref{Weierstrassfn}) rather than 1/12.} in Weierstrass form is:
\be
{d^2 h\over du^2}+\left[\kappa\left(1-{\nu(\nu+1)\over12}\right)-
\nu(\nu+1){\cal P}\left(u;\frac1{12},g_3\right)\right]h=0,
\label{Lame}
\ee
where ${\cal P}\left(u;g_2,g_3\right)$ is the Weierstrass elliptic function (see Figs. 1 and 2) defined by
\be
u=\int_{\infty}^{{\cal P}\left(u;g_2,g_3\right)}{dt\over\sqrt{4t^3-g_2t-g_3}}\ .
\label{Weierstrassfn}
\ee
The discriminant in the general case is 
\be 
\Delta\equiv g_2^3-27\,g_3^2,
\ee
which reduces to 
\be 
\Delta= {(2-b)b\over (12)^3}\, ,
\ee
for (\ref{Lame}) and is positive only for $0<b<2$. For these special models the cubic in
(\ref{Weierstrassfn}) has three real roots and for Big Bang cosmologies these are the 
recollapsing models. For other values of $b$ the cubic has only one real root and a pair of complex 
conjugates ones.

The Riemann surface for the integrand of (\ref{Weierstrassfn}) is double layered because of the 
square root and has three
branch points because of the cubic.  The two layers of the square root make ${\cal P}(u)={\cal P}(-u)$ 
and the branch 
points make ${\cal P}(u)$ doubly periodic. The integral is path dependent and by circumscribing two 
branch points  the value of the integral is increased by a fixed period.  By circumscribing another 
pair of branch 
points a second period is obtained. A third choice of branch points gives a combination 
of the first two periods.
 The domain for ${\cal P}(u)$ is usually given as a fundamental 
parallogram (cell) centered at the origin of the complex $u$ plane 
and generated by two half periods $\omega$ and 
$\omega^{\prime}$, see Fig. 1. What is needed to determine Luminosity-distance from (\ref{Lame}) are  
values for $u$ in this cell on the negative real axis, as determined by (\ref{u}). 
In general, ${\cal P}$ has only one singularity in a 
fundamental cell, a second order pole at $u=0$ with vanishing residue,
for which ${\cal P}(u)-1/u^2$ is not only analytic but vanishes at $u=0$.
To use (\ref{Lame}) to determine luminosity distance the boundary conditions (\ref{Aboundary}) 
on the area $A$ must be translate to boundary 
conditions on $h$:
\bea
h|_s&=&0, \nonumber\\
{dh\over du}\bigg|_s&=&-{c\over H_0\sqrt{|1-\OO|}}.
\label{hboundary}
\eea
The resulting luminosity distance is then given by
\be
D_{\ell}(z_s)=(1+z_s)h(u(0)).
\label{Dlh}
\ee
In Section \ref{special}, equation (\ref{Lame}) is integrated for some special cases.

\subsection{$\OO\equiv \OM+\OL=1$, Flat FLRW Models} \label{OO=1}

For the flat case the required change of variables for (\ref{area}) is 
\bea 
y&=&\OM\,(1+z),\nonumber\\ 
h&=&(1+z)\sqrt{{A\over\delta\Omega_s}},
\label{heunflatvariables} 
\eea
and the resulting equation in Heun form is 
\be
{d^2h\over dy^2} +{{3\over 2}y^2\over y^3+\OM^2\OL}\
{dh\over dy} - { {1\over 4}\nu(\nu+1)y\over y^3+\OM^2\OL}\ h = 0\ . 
\label{heunflat} 
\ee
This equation was described in \cite{KR98} but was never actually written down. However, 
solutions to (\ref{heunflat}), linear Heun functions,  were used to give an expression 
for luminosity distance for flat FLRW.
When $\OO=1$, Kantowski and Thomas \cite{KT} showed that (\ref{area}) was equivalent to the Legendre equation 
and
$D_{\ell}(z)$ was given in terms of associated Legendre functions, as well as equivalent 
hypergeometric functions. 
An equivalent hypergeometric result was independently given by Demianski et al. \cite{DM}. 
Sereno et al. \cite{SCPdeR},
\cite{SPS} have shown that the equation remains hypergeometric 
even when generic quintessence is introduced.
  
To transform (\ref{heunflat}) into Lame$^{\prime}$ form the change of dependent variable is 
the same as in (\ref{heunflatvariables});
however, the change of independent variable from $z$ to $u$ simplifies to
\be
u(z)\equiv\int_{\infty}^{p(z)}{dt\over\sqrt{4t^3+4}}= 
-{1\over \sqrt{p(z)}}\ {}_2F_1\left(\frac16,\frac12;\frac76,-{1\over p(z)^3}\right)< 0,
\label{uOO=1}
\ee
where
\be
p(z)\equiv \left({\OM\over 1-\OM}\right)^{1/3}(1+z)\ > 0.
\label{pz0}
\ee
The area equation for flat FLRW in Lame$^{\prime}$ form becomes:
\be
{d^2 h\over du^2}-\nu(\nu+1){\cal P}\left(u;0,-4\right)h=0,
\label{LameOO=1}
\ee
where 
\be
{d{\cal P}\over du}=\sqrt{4{\cal P}^3+4}={2\over \sqrt{1-\OM}}\ \sqrt{1+\OM[(1+z)^3-1]}
\ee
Boundary conditions (\ref{hboundary}) on $h$ change to:
\bea
h|_s&=&0,\nonumber\\
{dh\over du}\bigg|_s&=&-{2c\over H_0[\OM^2(1-\OM)]^{1/6}},
\label{hboundaryOO=1}
\eea
but the connection of $h$ to $D_{\ell}(z)$ is unchanged from (\ref{Dlh}).

\section{THREE SOLVABLE CASES} \label{special}
To give useful and/or interesting expressions for 
luminosity distance as a function of redshift requires a knowledge of the special 
functions associated with the defining differential equations (\ref{heun}) or (\ref{Lame}). In
Kantowski \cite{KR98} special functions associated with (\ref{heun}) called Heun functions were used to give $D_{\ell}(z)$; however, 
Lame$^{\prime}$ functions 
defined by the Lame$^{\prime}$ equation (\ref{Lame}) are not defined for arbitrary values of the constant 
$H$ (see footnote 1).
Most of the special functions defined by (\ref{Lame}) relate to boundary value problems where $H$ is 
restricted to discrete characteristic values, see \cite{Erdeliy}, \cite{Arscott}, and \cite{WW}.  
The characteristic values do not coincide with the
values needed in (\ref{Lame}). However, there are three values $\{0,1,2\}$ for $\nu$ for
which a complete solution to (\ref{Lame}) can be given. In \cite{KKT} it was recognized that
the solution of (\ref{area}) could be given as elliptic integrals in Legendre form for these three cases. It is 
for these same three cases that the complete solution to (\ref{Lame}) can be given and hence
for these three cases the luminosity distance can be given by additionally applying boundary conditions,
(\ref{hboundary}) or (\ref{hboundaryOO=1}), 
and using (\ref{Dlh}). 
Because the Weierstrass form of Lam\'e equation contains the Weierstrass elliptic function 
its solution will be given in terms of Weierstrass elliptic integrals.

\subsection{$\nu=0$, The Filled Beam Case, Standard FLRW} \label{nu=0}

The one case for which (\ref{Lame}) is easily solvable is $\nu=0$,
the homogeneous filled beam FLRW case. 
\subsubsection{$\nu=0,\ \OO\equiv \OM+\OL\ne 1$, The Non-Flat Case}

The general solution is:
\be
h(u)=c_1\ S_{\kappa}[u-c_0],
\label{h-sol}
\ee
where $c_0$ and $c_1$ are integration constants and $S_{\kappa}[\ ]$ is 
one of two functions,
\[
S_{\kappa}[\ ]= \left\{
\begin{array}{l c  l}
{\rm sinh}[\ ] &:& \kappa= -1, \\
{\rm sin}[\ ] &:& \kappa = +1.
\end{array}
\right.
\]
Eqn. (\ref{h-sol}) results in an expression for luminosity distance 
\be
D_{\ell}(\OM,\OL,\nu=0;z)={c(1+z)\over H_0\sqrt{|1-\OO|}}S_{\kappa}\left[ u(z)-u(0)\right],
\label{DlSchucking}
\ee
where the  source redshift has been simplified from $z_s$ to $z$ and $u(z)$ is defined by (\ref{u}).
If 
${\cal P}^{-1}\left(p\,;g_2,g_3\right)$, the inverse of the Weierstrass function
defined in (\ref{Weierstrassfn}),
\be
{\cal P}^{-1}\left(p\,;g_2,g_3\right)\equiv\int_{\infty}^p{dt\over\sqrt{4t^3-g_2t-g_3}},
\label{WeierstrassInv}
\ee
is used to evaluate $u(z)$
the two inverse Weierstrass functions in (\ref{DlSchucking})
 can be combined using the identity:
\be
{\cal P}(u+v)=\frac14\left[{ {\cal P}^{\prime}(u)-{\cal P}^{\prime}(v)\over {\cal P}(u)-{\cal P}(v)}\right]^2
-{\cal P}(u)-{\cal P}(v),
\label{P-identity}
\ee
to give
\bea
&&u(z)-u(0)=
\left[{\cal P}^{-1}\left(p(z);\frac1{12},g_3\right)-{\cal P}^{-1}\left(p(0);\frac1{12},g_3\right)\right]
\label{u-u}\\
&&=
-{\cal P}^{-1}\left( {\kappa\over 6}+{\left[\sqrt{(1+z)^2(1+\OM z)-z(z+2)\OL}+1\right]^2-\OM(z+2)z^2
\over 4|1-\OO|z^2}
;\frac1{12},g_3\right)>0.
\nonumber
\eea
The resulting expression for (\ref{DlSchucking}) was first given
by Kaufman Schucking \cite{KSSE} and \cite{KS}.\footnote{
A factor of 1/2 was left out of (4) of \cite{KS} compared to (32) of \cite{KSSE}
and compared to (\ref{DlSchucking}) combined with (\ref{u-u}) above.}
Care has to be exercised when using the identity (\ref{u-u}) in (\ref{DlSchucking}). 
The inverse function  (\ref{WeierstrassInv}) is multi-valued (\eg its value ($\pm$)
depends on $p$ 's location on the Riemann surface of the square root appearing in (\ref{WeierstrassInv})). 
In \cite{KKT}, (\ref{DlSchucking}) 
was given in terms of incomplete Legendre 
elliptic integral of the first kind, $F(\phi,{\rm k})$, rather than inverse Weierstrass
elliptic integral ${\cal P}^{-1}(p\,;g_2,g_3)$.
\subsubsection{$\nu=0,\  \OO\equiv \OM+\OL=1$, Standard FLRW, The Filled-Beam Flat Case}
For the flat case the general solution to (\ref{LameOO=1}) is 
\be
h(u)=c_1\,(u-c_0),
\label{h-solOO=1}
\ee
which results in a luminosity-distance
\be
D_{\ell}(\OM,\OL=1-\OM,\nu=0;z)={2c(1+z)\over H_0[\OM^2(1-\OM)]^{1/6}}\biggl( u(z)-u(0)\biggr).
\label{DlSchuckingOO=1}
\ee
The $z$ dependence of $u(z)-u(0)$ can be evaluated using (\ref{WeierstrassInv})
which is equivalent to a hypergeometric function by (\ref{uOO=1}),
\bea
u(z)&=&{\cal P}^{-1}\biggl(p(z);0,-4\biggr)\nonumber\\
&=&-{1\over \sqrt{p(z)}}\ {}_2F_1\left(\frac16,\frac12;\frac76,-{1\over p(z)^3}\right)\nonumber\\
&=& -{1\over p(z)^{1/4}}\ 2^{1/6}\ \Gamma\left(\frac76\right)P^{-1/6}_{-1/6}\left(\sqrt{1+{1\over p(z)^{3}}}\right)< 0,
\label{uzOO=1}
\eea
where $p(z)$ is given in (\ref{pz0}).
The third equality is simply a relation that exists between all associated 
Legendre functions and some hypergeometric functions.
The addition law (\ref{P-identity}) results in the simplification
\bea
u(z)-u(0)&=& 
{\cal P}^{-1}\left(p(z);0,-4\right)-{\cal P}^{-1}\left(p(0);0,-4\right)
\nonumber\\
&=& 
-{\cal P}^{-1}\left(
\left[{2\sqrt{1+\OM z(3+3z+z^2)}+2+\OM z(3+z)\over [\OM^2(1-\OM)]^{1/3} z^2}\right]
;0,-4\right)
> 0.
\label{u-uOO=1}
\eea
The expression obtained by using the last 
equality in (\ref{u-uOO=1}) results in an expression obtained by \cite{KS} when the identity 
${\cal P}\left(u;g_2,g_3\right)$=$T^2{\cal P}\left(Tu;T^{-4}g_2,T^{-6}g_3\right)$ 
is used.
When the equivalence of the inverse Weierstrass function ${\cal P}^{-1}(p;0,-4)$ and 
the hypergeometric function ${}_2F_1$ as implied by (\ref{uzOO=1}) is used in (\ref{DlSchuckingOO=1}) 
a new and simplified  expression 
for luminosity distance for standard (filled beam) flat FLRW results:
\bea
D_{\ell}(\OM,\OL&=&1-\OM,\nu=0;z)=
{ 
2\,c\,(1+z)\,z 
\over 
\left[2\sqrt{1+\OM z(3+3z+z^2)}+2+\OM z(3+z)\right]^{1/2}
}\ \times
\nonumber\\
&&{}_2F_1\left(\frac16,\frac12;\frac76,
- \left[{ [\OM^2(1-\OM)]^{1/3} z^2 \over 2\sqrt{1+\OM z(3+3z+z^2)}+2+\OM z(3+z)} \right]^3
\right)\nonumber\\
&&=
{ 
c\ 2^{7/6}\ \Gamma\left(\frac76\right)\ (1+z)\ z 
\over 
\left[2\sqrt{1+\OM z(3+3z+z^2)}+2+\OM z(3+z)\right]^{1/4}
}\ \times
\nonumber\\
&&P^{-1/6}_{-1/6}\left(
\sqrt{
1+\left[{ [\OM^2(1-\OM)]^{1/3} z^2 \over 2\sqrt{1+\OM z(3+3z+z^2)}+2+\OM z(3+z)} \right]^3
}
\right)
.
\label{Dl_nu0FlatHyper}
\eea
See \cite{KKT}, \cite{KT}, and  \cite{DM} for previously known expressions.
The first expression in (\ref{Dl_nu0FlatHyper}) should be quite efficient for numerical evaluation because 
${}_2F_1=1+{\cal O}(z^6)$ and the argument of the hypergeometric function remains close to zero
for reasonable $z$. The second expression involving the associated Legendre function $P^{-1/6}_{-1/6}$ 
seem less useful but is included for completeness.

 \subsection{$\nu=1$, The 66\% Filled Beam FLRW} \label{nu=1}

This is a partially filled beam case corresponding to having $\nu(1+\nu)/6=1/3$ of the mass density 
excluded from the observing beam. For this case (\ref{Lame}) can be integrated in a straight forward manner giving
the general solution in terms of two integration constants $c_0$ and $c_1$:
\be
h=c_1\sqrt{{\kappa 5\over 6}+{\cal P}\left(u;\frac1{12},g_3\right)}
S_b\biggl[\sqrt{\left|{486-b\over 3^32^5}\right|}\int^u_{c_0}
{du\over 
\left[{\kappa 5\over 6}+{\cal P}\left(u;\frac1{12},g_3\right)\right]}\biggr]
\label{h1},
\ee 
where 

\[
S_{b}[\ ]= \left\{
\begin{array}{l c  l}

{\rm sinh}[\ ] &:& b<0\, \\
{\rm sin}[\ ] &:& 0<b<486, \\
{\rm sinh}[\ ] &:& 486<b.
\end{array}
\right.
\]
Special cases like $b=486$ were dealt with in \cite{KKT} but are simply neglected here.
The above result can be checked using the identities 
\bea
{\cal P}^{\prime\prime} &=& 6{\cal P}^{2}-\frac12g_2,\nonumber\\
({\cal P}^{\prime})^2 &=& 4{\cal P}^{3}-g_2{\cal P}-g_3
\label{identity}
\eea
satisfied by all by all Weierstrass ${\cal P}$ functions.
The denominator satisfies:
\be
{\kappa 5\over 6}+{\cal P}\left(u;\frac1{12},g_3\right)>{ \kappa3\over4}
\label{denominator}\ee
and can cause complications for negative curvature models if it vanishes along 
the path of integration. 
 When boundary conditions  (\ref{hboundary}) are applied, the luminosity distance becomes:   

\bea
D_{\ell}(\OM,\OL,\nu=1;z)&=&{c\over H_0} (1+z)
\sqrt{
\left| 
{
\left[3-\OM(1+z)/(1-\OO)\right]\left[3-\OM/(1-\OO)\right]
\over 
(1-\OO)(486-b)/(3^32)
}
\right|
}
\times
\nonumber\\
&&
{\rm Sign}\left[{\kappa5\over 6}+p(0)\right] S_{(\OM,\OL,z)}\biggl[ 
\sqrt{\left|{486-b\over 3^32^5}\right|}\int^{u(z)}_{u(0)}{du\over 
\left[{\kappa5\over 6}+{\cal P}\left(u;\frac1{12},g_3\right)\right]}
\biggr],
\label{D_nu1}
\eea
where $u(z)$ is given by (\ref{u}) and
\[
S_{(\OM,\OL,z)}[\ ]= \left\{
\begin{array}{l c  l}
{\rm cosh}[P\ ] &:& b<0\,\&\,\left[3-\OM(1+z)/(1-\OO)\right]\left[3-\OM/(1-\OO)\right]<0, \\
{\rm sinh}[\ ] &:& b<0\,\&\,\left[3-\OM(1+z)/(1-\OO)\right]\left[3-\OM/(1-\OO)\right]>0, \\
{\rm sin}[\ ] &:& 0<b<486, \\
{\rm sinh}[\ ] &:& 486<b.
\end{array}
\right.
\] 

For the first case, $b<0$, the principal value of the integral, P$\int$, in (\ref{D_nu1}) is needed 
when the denominator of the integrand 
vanishes between the source and observer, see (\ref{denominator}). In this case the argument of hyperbolic trig 
function sinh[\,] develops an imaginary part, $i\pi/2$, which 
turns it into a cosh[\,] of the principal part. 

The integral in (\ref{D_nu1}) is an elliptic integral 
written in terms of the Weierstrass ${\cal P}$ function and as such can always be
evaluated in terms of the three Weierstrass functions ${\cal P}^{-1}, \zeta,$ 
and ${\sigma}$ (see 13.14 of \cite{Erdeliy}). 
For (\ref{D_nu1}) the integrand is an elliptic function with the same fundamental cell as 
${\cal P}(u;1/12,g_3)$ and has two first order poles at $u=\beta_{\pm},\ (\beta_-=-\beta_+)$ 
within the cell shown in Figs. 1 and 2.  The points $\beta_{\pm}$ are those two 
lattice points defined by 
\be
\beta={\cal P}^{-1}(-\kappa5/6;1/12,g_3)
\ee
and which lie within the fundamental cell.
An elliptic function is uniquely given (up to a constant that can easily be evaluated)
by the principal parts of its Laurent expansions about its poles within a fundamental cell.      
Because the principal parts can be expressed in terms of the 
Weierstrass $\zeta$-function and its derivatives, a unique expansion of the elliptic function
exists in terms of these functions. 
The expansion for the integrand in (\ref{D_nu1}) is:
\be
{1\over 
\left[{\kappa 5\over 6}+{\cal P}\left(u;\frac1{12},g_3\right)\right]
}={1\over {\cal P}^{\prime}\left(\beta_+;\frac1{12},g_3\right)}
\left[2\,\zeta\left(\beta_+;\frac1{12},g_3\right)+
\zeta\left(u-\beta_+;\frac1{12},g_3\right)-\zeta\left(u+\beta_+;\frac1{12},g_3\right)\right].
\label{integrand_nu1} 
\ee 
In (\ref{integrand_nu1}) the first two terms are constants and the second two are the principal parts
at the two first order poles in the fundamental cell. The value of 
\be
{\cal P}^{\prime}\left(\beta_+;\frac1{12},g_3\right)={}_{\pm}\sqrt{{-\kappa(486-b)\over 6^3}}
\ee
is real except when $0<b<486$.
When integrated (\ref{integrand_nu1}) gives
\bea
\int^{u(z)}_{u(0)}{du\over 
\left[{\kappa 5\over 6}+{\cal P}\left(u;\frac1{12},g_3\right)\right]
}
&=& {1\over {\cal P}^{\prime}\left(\beta_+;\frac1{12},g_3\right)}
\Biggl\{2\,\zeta\left(\beta_+;\frac1{12},g_3\right)  [u(z)-u(0)] 
\nonumber\\   
&+&\log\left[ 
{
\sigma\left(u(z)-\beta_+;\frac1{12},g_3\right)\sigma\left(u(0)+\beta_+;\frac1{12},g_3\right)
\over
\sigma\left(u(0)-\beta_+;\frac1{12},g_3\right)\sigma\left(u(z)+\beta_+;\frac1{12},g_3\right)
}\right]
\Biggr\}
\label{D_nu1b}
\eea
The simplification of (\ref{D_nu1b}) similar to the simplification of (\ref{D_nu2b}) must exist but
escapes the author. To evaluate (\ref{D_nu1b}) and hence (\ref{D_nu1}) the three quantities $u(z), u(0),$ and $u(z)-u(0)$
can be evaluated as in (\ref{u-u}) combined with (\ref{p}). Fortran routines for $\zeta$ and $\sigma$ are not available
at this time, 
however, they do exist in Mathematica.
In \cite{KKT} the area equation (\ref{area}) was integrated directly giving luminosity-distance in terms of 
Legendre elliptic integrals $F(\phi,{\rm k})$ and 
$\Pi(\phi,{\alpha}^2,{\rm k})$.  Fortran routines for these are available (see the Numerical Recipes website).

\subsection{$\nu=2$, The Empty Beam Case} \label{sectnu=2}

This is the case where all material is exterior to the observation beam and its effect through the
Weyl curvature (lensing) is neglected. For this case (\ref{Lame}) can be integrated in a straight 
forward manner giving
the general solution in terms of two integration constants $c_0$ and $c_1$,
\be
h=c_1\left[{\kappa\over 12}+{\cal P}\left(u;\frac1{12},g_3\right)\right]\int^u_{c_0}{du\over 
\left[{\kappa\over 12}+{\cal P}\left(u;\frac1{12},g_3\right)\right]^2}.
\label{hP}
\ee
As with the $\nu=1$ case, this result can be checked by using (\ref{identity}).
When boundary conditions (\ref{hboundary}) are applied the luminosity distance becomes:
\be
D_{\ell}(\OM,\OL,\nu=2;z)={c\over H_0}\left[{\OM\over 4|1-\OO|}\right]^2{ (1+z)^2\over \sqrt{|1-\OO|}}
\int^{u(z)}_{u(0)}{du\over 
\left[{\kappa\over 12}+{\cal P}\left(u;\frac1{12},g_3\right)\right]^2},
\label{D_nu2}
\ee
where $u(z)$ is given by (\ref{u}). 
For Big Bang models the denominator in the integral doesn't vanish for any $z \ge 0$ and hence problems that 
occurred for the $\nu=1$ case do not appear for $\nu=2$.
Just as with (\ref{D_nu1}) the integral in (\ref{hP}) is an elliptic integral 
written in terms of the Weierstrass ${\cal P}$ functions and as such can also be
evaluated in terms of the three Weierstrass function ${\cal P}^{-1}, \zeta,$ 
and ${\sigma}$. 
For (\ref{D_nu2}) the integrand is again an elliptic function with the same fundamental cell as 
${\cal P}(u;1/12,g_3)$ but has two second order poles at $u=\beta_{\pm},\ (\beta_-=-\beta_+)$ 
within this cell.  The two poles $\beta_{\pm}$ now belong to the
lattice defined by 
\be
\beta={\cal P}^{-1}(-\kappa/12;1/12,g_3).
\ee
The unique expansion for the integrand in (\ref{D_nu2}) is:
\bea
{1\over 
\left[{\kappa\over 12}+{\cal P}\left(u;\frac1{12},g_3\right)\right]^2}
&=& 
{1\over \left[{\cal P}^{\prime}(\beta_+)\right]^2}
\biggl\{\ \left[\zeta^{\prime}\left(-\beta_+;1/12,g_3\right)
+ \zeta^{\prime}\left(\beta_+;1/12,g_3\right)\right]\nonumber\\
&-& 
\left[\zeta^{\prime}\left(u-\beta_+;1/12,g_3\right)+ 
\zeta^{\prime}\left(u+\beta_+;1/12,g_3\right)\right]\biggr\},
\label{integrand_nu2}
\eea
where $\zeta^{\prime}(u)\equiv -{\cal P}(u)$ is the first derivative of the Weierstrass $\zeta$-function.
In (\ref{integrand_nu2}) the first two terms are constants and the second two are the principal parts
at the two second order poles.

This expression can be easily integrated to give a value for (\ref{D_nu2}),
\bea
D_{\ell}(\OM,\OL,\nu=2;z)&=&{c\over H_0}{\sqrt{|1-\OO|}\over \OL}(1+z)^2
\biggl\{
{\kappa\over 6}\
\left[u(z)-u(0)\right]
\nonumber\\
&-& \left[{\cal \zeta}\left(u(z)-\beta_+;\frac1{12},g_3\right)-{\cal \zeta}\left(u(0)-\beta_+;\frac1{12},g_3\right)\right]
\nonumber\\
&-&\left[{\cal \zeta}\left(u(z)+\beta_+;\frac1{12},g_3\right)-{\cal \zeta}\left(u(0)+\beta_+;\frac1{12},g_3\right)\right]
\biggr\}.
\eea
The  $\zeta$ terms 
can be combined using the $\zeta$-identity:
\be
\zeta(u+v)=\zeta(u)+\zeta(v)+\frac12\left[{ {\cal P}^{\prime}(u)-{\cal P}^{\prime}(v)\over {\cal P}(u)-{\cal P}(v)}\right],
\label{zeta-identity}
\ee
along with (\ref{P-identity}).
Using ${\cal P}(\beta_+)=-\kappa/12$ and ${\cal P}^{\prime}(\beta_+)=\pm \sqrt{\kappa\,b/(3^32^3)}=\pm\OM\sqrt{\OL}/(4|1-\OO|^{3/2})$
the resulting luminosity-distance becomes 
\bea
&&D_{\ell}(\OM,\OL,\nu=2;z)={c\over H_0}{\sqrt{|1-\OO|}\over \OL}(1+z)^2
\Biggl[
{\kappa\over 6}\ \bigl[u(z)-u(0)\bigr]
-2\,\zeta\biggl(u(z)-u(0);\frac1{12},g_3\biggr)
\nonumber\\
&&
+\Biggl\{
{\kappa\over 2}+
{1\over 4|1-\OO|}
\biggl[ 2\left({1\over z^2}+{\sqrt{\OL}\over (1+z)^2}\right)\sqrt{(1+z)^2(1+\OM z)-z(z+2)\OL}
\nonumber\\
&& 
+\left(3+{2\over z}+{2\over z^2}\right)+\OM\left({1\over z}-2\right)-\OL\left({2\over z}-{2\over (1+z)^2}-1\right)+2\sqrt{\OL}
\biggr]
\Biggr\}^{1/2}
\nonumber\\
&& 
+\Biggl\{
{\kappa\over 2}+
{1\over 4|1-\OO|}
\biggl[ 2\left({1\over z^2}-{\sqrt{\OL}\over (1+z)^2}\right)\sqrt{(1+z)^2(1+\OM z)-z(z+2)\OL}
\nonumber\\
&& 
+\left(3+{2\over z}+{2\over z^2}\right)+\OM\left({1\over z}-2\right)-\OL\left({2\over z}-{2\over (1+z)^2}-1\right)-2\sqrt{\OL}
\biggr]
\Biggr\}^{1/2}
\Biggr]
\nonumber
\\
\label{D_nu2b}
\eea
The $u(z)-u(0)$ arguments are given in (\ref{u-u}) resulting in a distance-redshift expressed in terms of
the two independent Weierstrass elliptic integrals ${\cal P}^{-1}$ and $\zeta$.

If the integration in (\ref{D_nu2}) is changed from $u$ to $z$ by using (\ref{u}) and (\ref{p}),
the luminosity distance becomes the familiar elliptic integral,
\bea 
D_{\ell}(\OM,\OL,\nu=2;z)&=&{c\over H_0}(1+z)^2
\int_0^z{dz\over(1+z)^2\sqrt{(1+z)^2(1+\OM z)-z(z+2)\OL}},
\label{nu=2}
\eea
obtainable directly from (\ref{area}).
In \cite{KKT} the value of this integral was given in terms of Legendre elliptic integrals $F(\phi,{\rm k})$ and 
$E(\phi,{\rm k})$ in a somewhat more complicated expression. 

\section{Conclusions} \label{conclusions}

In this paper the partially filled beam Hubble relation for pressure free FLRW cosmologies has been shown to 
satisfy the Lame$^{\prime}$ equation (\ref{Lame}) or (\ref{LameOO=1}) 
when appropriate dependent and independent variables
are chosen. 
For three values of the beam's filling parameter ($\nu=0,1,2$), general solutions 
to the Lame$^{\prime}$ equation were given in terms of Weierstrass elliptic integrals. These 
integrals were used to give the Hubble relation for these three cases. Practical use of these expressions
awaits implementation of fast fortran routines for the Weierstrass integrals. The results
given here were checked using the slow routines of Mathematica. The  Lame$^{\prime}$ equation 
has not found its way into the physics or astronomy literature much in the past, however, 
recently some  use in string theory is showing up, \cite{GKLS}.

The partially filled beam Hubble relation for flat FLRW 
is known to satisfy the associated Legendre equation \cite{KT} and hence 
can be written as a linear combination of two associated Legendre functions. 
In general the two coefficients of these functions both depend on associated 
Legendre functions bringing the total to four. For the filled  beam case,
two of the functions reduce to ordinary functions leaving two to be evaluated
as associated Legendre or as their equivalent hypergeometric functions. For this special case
an addition formula for the Weierstrass ${\cal P}$ functions allowed the discovery of 
a  Hubble relation (\ref{Dl_nu0FlatHyper}) containing only one associated Legendre or equivalent 
hypergeometric function. A fortran implementation of this expression should rapidly converge 
(see http://www.nhn.ou.edu/$\sim$thomas/z2dl.html for earlier code).

The Hubble relations given here are good for observations made on distant objects 
when lensing by inhomogeneous mass external to the observing beam can be neglected and
when transparent mass interior to the beam is scattered along the beam and not concentrated
at isolated regions. For cases where mass concentrations external and/or internal are important,
expressions given here are necessary to compute source-deflector, source-observer, etc., distances
for gravitational lensing, see \cite{BK} and \cite{CK}. 
In \cite{KT} it was argued that even though the filling parameter $\nu$ could modify observed magnitudes 
by small amounts, $\sim 0.05$, a tenfold
increase the number of SNe\,Ia $m(z)$ data ($\sim 600$) would statistically necessitate limiting its value.
This amount of data is to become available from the SuperNova Acceleration Probe
(SNAP - http://snap.lbl.gov).

The $m(z)$ relations given here can also be used to help 
understand ray tracing simulations that predict weak lensing amplifications distributions 
for SNe, e.g.,  \cite{WJ2}, \cite{PP2}, \cite{HW}, and \cite{BGGM}.
The minimum amplification for weak lensing occurs for beams 
that have a minimum amount of transparent and homogeneous mass within them and for which inhomogeneities are 
far enough away to have negligible lensing effect
(see Fig. 3 for a generic histogram of lensing amplifications, \cite{WHM}). 
Some work has been done to find an effective filling fraction 
$\alpha\equiv {\rho_H/\rho_m}={(3+\nu)(2-\nu)/ 6}$, \cite{ME}, \cite{TPN}.
Such a filling parameter $\nu_{eff}$ is not necessarily
available a priori for the simulation; however,  it
can be obtained by comparison of the simulation's histogram with 
the partially filled beam de-amplifications given here, 
\be
Amp_{min} =D_{\ell}^2(\OM,\OL,\nu=0;z)/D_{\ell}^2(\OM,\OL,\nu_{eff};z).
\label{amp-min}
\ee 
The ratio of the luminosity with an effective filling parameter, to the average luminosity
can be taken to be the minimum amplification of the simulation. 

\begin{acknowledgments}
The author thanks R. C. Thomas for making suggestions to improve this manuscript.
\end{acknowledgments}

\clearpage

\begin{figure}
\includegraphics{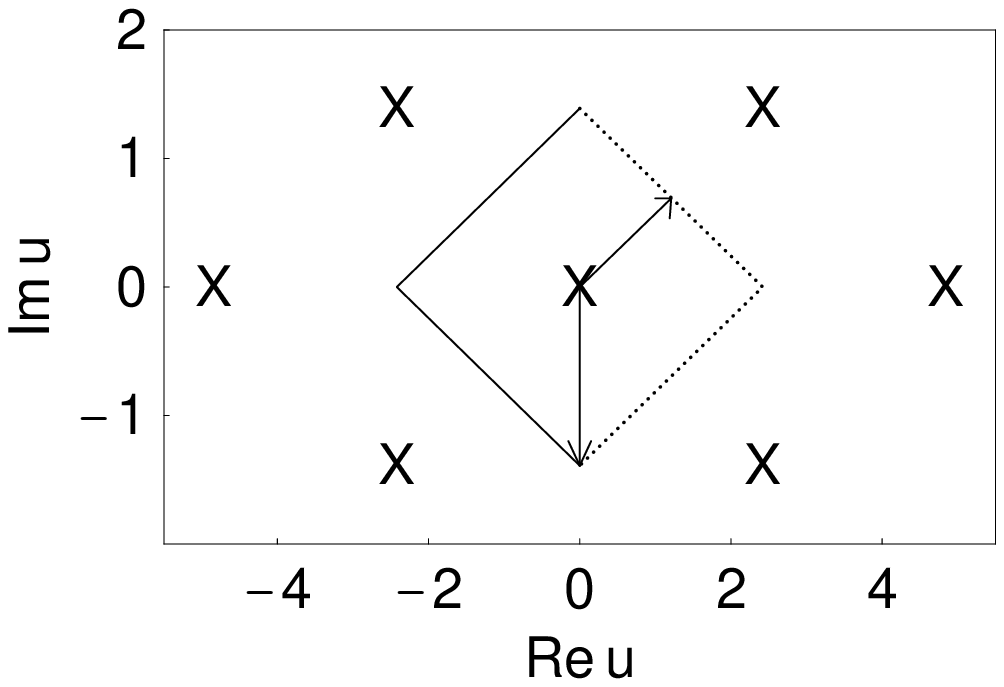}
\caption[fig1_3.eps]{A typical fundamental cell (domain) for the doubly periodic Weierstrass elliptic  
function ${\cal P}\left(u;1/12,g_3=-1.75\right)$. 
The cell and half periods, $\omega=-1.39\, i$ and $\omega^{\prime}=1.21+0.69\, i$, are for a universe with
$\OM=0.2$ and $\OL=0.7$. The periodic locations of the $2^{nd}$-order poles are indicated by the ${\bf X}$'s.
\label{fig1_3}}
\end{figure}

\begin{figure}
\includegraphics{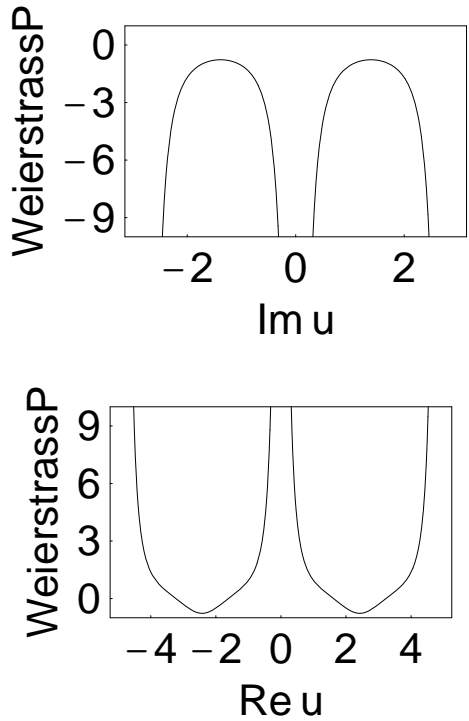}
\caption[fig2_3.eps]{Plots of ${\cal P}\left(u;1/12,g_3=-1.75\right)$ along the real and imaginary 
u-axies for  a universe with $\OM=0.2$ and $\OL=0.7$. The function is real on these lines.
\label{fig2_3}}
\end{figure}

\begin{figure}
\includegraphics{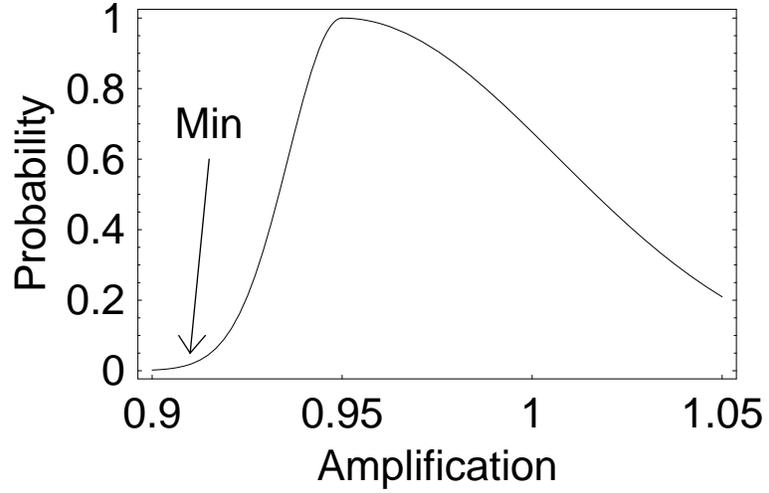}
\caption[fig3_3.eps]{A schematic distribution of weak lensing amplifications (relative to a homogeneous
universe) 
expected for observing a set of SNe at a fixed redshift. The indicated minimum can be used to determine an
effective value of the filling parameter $\nu$ by using (\ref{amp-min}).
\label{fig3_3}}
\end{figure}

\end{document}